# Arbitrary manipulations of dual harmonics and their wave behaviors based on space-time-coding digital metasurface

Jun Yan Dai,[1,2] Jin Yang,[1] Wankai Tang,[3] Ming Zheng Chen,[1] Jun Chen Ke,[1] Qiang Cheng,[1, a] Shi Jin,[3, a] and Tie Jun Cui[1, a]

[1] State Key Laboratory of Millimeter Waves, Southeast University, Nanjing, 210096, China

[2] State Key Laboratory of Terahertz and Millimeter Waves, Department of Electronic Engineering, City University of Hong Kong, 999077, Hong Kong

[3] National Mobile Communications Research Laboratory, Southeast University, Nanjing, 210096, China

[a] Authors to whom correspondence should be addressed: qiangcheng@seu.edu.cn, jinshi@seu.edu.cn, and tjcui@seu.edu.cn

## Abstract

Space-time modulated metasurfaces have attracted significant attention due to the additional degree of freedom in manipulating the electromagnetic (EM) waves in both space and time domains. However, the existing techniques have limited wave control capabilities, leading to just a few feasible schemes like regulation of only one specific harmonic. Here, we propose to realize independent manipulations of arbitrarily dual harmonics and their wave behaviors using a space-time-coding (STC) digital metasurface. By employing different STC sequences to the reflection phase of the metasurface, independent phase-pattern configurations of two desired harmonics can be achieved simultaneously, which further leads to independent beam shaping at the two harmonic frequencies. An analytical theory is developed to offer the physical insights in the arbitrary dual-harmonic manipulations of spectra and spatial beams, which is verified by experiments with good agreements. The presented STC strategy provides a new way to design multifunctional programmable systems, which will find potential applications such as cognitive radar and multi-user wireless communications.

# I. INTRODUCTION

Over the past decades, metamaterials comprising artificially engineered metallic or dielectric structures have attracted great attention owing to their customizable properties beyond the limitations of natural materials. The metamaterials have been widely investigated to provide unconventional ways to regulate the wave-matter interaction over subwavelength scale, leading to a myriad of exotic phenomena and promising applications [1-4]. As the two-dimensional (2D) version of metamaterials, metasurfaces hold the advantages of low loss, light weight, and easy integration in addition to the powerful control capabilities of electromagnetic (EM) waves. With the introduction of abrupt phase shift on the interface, the reflection and refraction of EM waves can be controlled with high efficiency, thus putting forward various interesting functions and novel devices in microwave [5-7], terahertz [8-10] and light regimes [11-13]. To simplify the design and optimization procedures, Cui *et al.* proposed a new strategy of coding metasurface, which is composed of digital coding meta-atoms and manipulates EM waves by coding sequence [14, 15]. With the integration of various active devices, e.g. PIN diodes and varactor diodes, the coding metasurface provides capabilities to dynamically switch the digital coding states, leading to the digital version of metasurface. The digital coding metasurface paves effective ways to perform amplitude/phase modulations (AM/PM) with the aid of digital signal processors (e.g. field-programmable gate array (FPGA) and microprogrammed control unit (MCU)), thus upgrading the metasurface to field-programmable and further building a bridge between the physical and digital worlds [16-21].

Recently, the emergence of time-varying metasurface has extended the research to time dimension by modulating the EM characteristics in the time domain [22-32]. The newly arisen approach has attracted increasing interests due to an additional degree of freedom to manipulate the EM waves, which brings peculiar phenomena and potential applications. In addition, with the combination of space modulation method, the space-time-modulated metasurface was also investigated through controlling the physical properties in both space and time domains, which apparently owns great values in providing entirely new physics and revolutionary technologies [33-42]. Many related studies have shown exciting functionalities such as breaking time-reversal symmetry and Lorentz reciprocity [22, 34, 41], frequency translation [25, 27, 38], pseudo Doppler effect [26], harmonic manipulation [24, 28, 35-37, 40-42], optical isolation [22], and direct information modulation [28-32]. In particular, as one of the few practical realizations, space-time-coding (STC) digital metasurfaces were proposed to facilitate the control of EM waves in both space and frequency domains, demonstrating the harmonic generation [28, 40], wavefront engineering [40-42], scattering signature manipulation [40], and programmable nonreciprocal effect [41]. However, the currently existing active metasurfaces lack efficient capabilities for EM-wave regulations, such as the amplitude-phase joint modulation. Therefore, despite that some ingenious methods have been proposed to improve the accuracy and reduce the complexity, most of the aforementioned time-varying or space-time-modulated metasurfaces can only focus on realizing flexible control of one specific harmonic or limited regulation of multiple harmonics, which motivates us to explore the implementation of arbitrary dual-harmonic manipulations.

Here, we propose a new STC strategy to realize arbitrarily independent manipulations of dual harmonics and their spatial beams with the STC digital metasurface. By employing different STC sequences to the reflection phase of the metasurface, independent phase-pattern configurations of two desired harmonics can be achieved simultaneously, which further leads to independent beam shaping at the two harmonic frequencies without affecting the original power distribution of harmonics. We present a theoretical strategy of phase syntheses for arbitrary dual harmonics and introduce an analysis method of scattering patterns for wavefront shaping. Two illustrative examples are provided to demonstrate the capability of the proposed approach for independent controls of dual harmonics. The first one aims to realize orbital-angular-momentum (OAM) generation and beam steering at the $+1^{st}$ and $+2^{nd}$-order harmonics, respectively; while the second example focuses on implementing different scattering patterns at the $+1^{st}$ and $-1^{st}$-order harmonics, respectively. We design and fabricate a 16 × 8 STC metasurface sample loaded with varactor diodes as prototype system for experimental validation, showing good match to the theoretical results. The proposed theory and method will improve the potential of the STC digital metasurfaces in the applications of imaging, radar, and wireless communication systems.

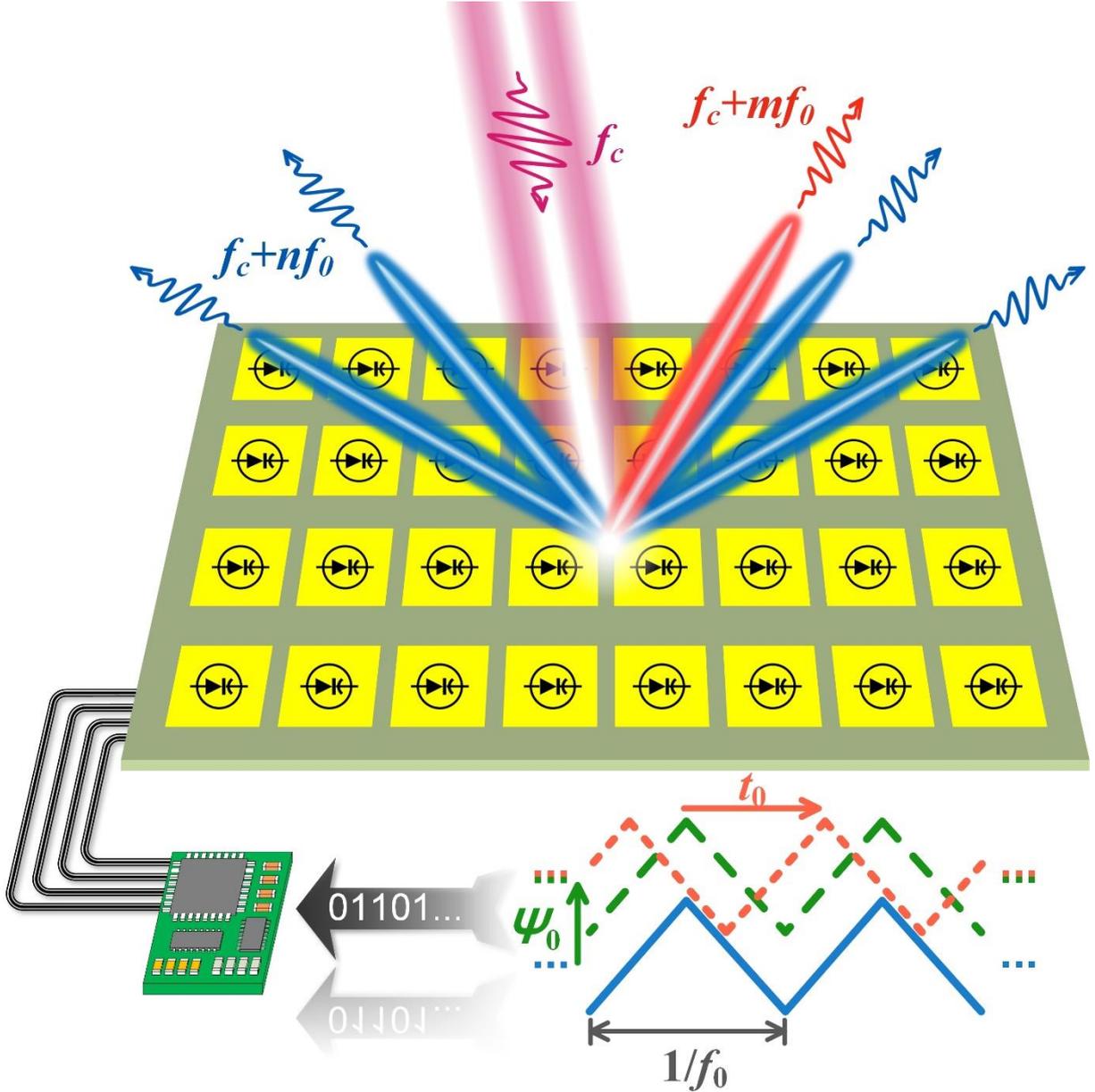

**FIG. 1.** Schematic description of the STC digital metasurface, which can manipulate beam shaping of arbitrary dual harmonics independently in programmable way by applying different STC sequences.

## II. THEORY AND METHODOLOGY

**Phase synthesis method of arbitrary dual harmonics.** We consider a reflective STC digital metasurface composing of $M \times N$ meta-atoms, as shown in Fig. 1, each of which is loaded with varactor diodes so that the reflection response can be controlled in a programmed way via FPGA. When the metasurface is normally illuminated by a monochromatic wave $E_i(t) = \exp(j2\pi f_c t)$, the reflected wave can be expressed as [28]

$$E_r(t) = E_i(t) \times \Gamma(t), \quad (1)$$

where $\Gamma(t) = A(t)\exp[j\psi(t)]$ is the periodic time-varying reflection coefficient of the metasurface, with $A(t)$ and $\psi(t)$ standing for the amplitude and phase, respectively. We remark that Eq. (1) is approximately valid when the time-varying frequency is much lower than the EM wave frequency $f_c$ [38, 43]. By applying the Fourier transform to Eq. (1), the expression of the reflected wave in the frequency domain reads as

$$E_r(f) = E_i(f) * \Gamma(f) = \delta(f - f_c) * \Gamma(f) = \Gamma(f - f_c), \quad (2)$$

in which $\delta(f - f_c)$ denotes the Dirac delta function at $f = f_c$. The periodic function $\Gamma(t)$ can be further decomposed into a linear combination of harmonically related complex exponentials $\Gamma(t) = \sum_{k=-\infty}^{\infty} a_k \exp\left(jk\frac{2\pi}{T}t\right)$, i.e. the Fourier series, whence $\Gamma(f) = \sum_{k=-\infty}^{\infty} a_k \delta(f - kf_0)$, where $f_0 = 1/T$ refers to the harmonic frequency determined by the function period T, and $a_k$ is the coefficient of the $k^{th}$-order harmonic component. Therefore, Eq. (2) finally becomes

$$E_r(f) = \sum_{k=-\infty}^{\infty} a_k \delta(f - kf_0 - f_c). \quad (3)$$

The new equation indicates that the spectrum of the reflected wave is composed of a series of harmonics around the central frequency $f_c$. According to the Fourier transform theory, the harmonic coefficient is given by

$$a_k = \frac{1}{T}\int_0^T \Gamma(t)\exp(-jk2\pi f_0 t)dt = A_k \exp(j\Psi_k), \quad (4)$$

where $A_k$ and $\Psi_k$ respectively represent the amplitude and phase of the $k^{th}$-order harmonic wave at frequency $f_c + kf_0$. Theoretically speaking, the amplitude and phase distributions of arbitrary harmonic can be achieved with meticulously designed $\Gamma(t)$. However, the strongly coupled amplitude and phase responses of the metasurface will constrain the feasibility of $\Gamma(t)$ waveform, resulting in many difficulties in implementing arbitrary harmonic distributions in practice. Refs. [40, 42] introduced some feasible algorithms to realize flexible harmonic controls. Although they are capable of modulating both phase and amplitude at either the central frequency or harmonic frequency, they are still limited to only one specific harmonic.

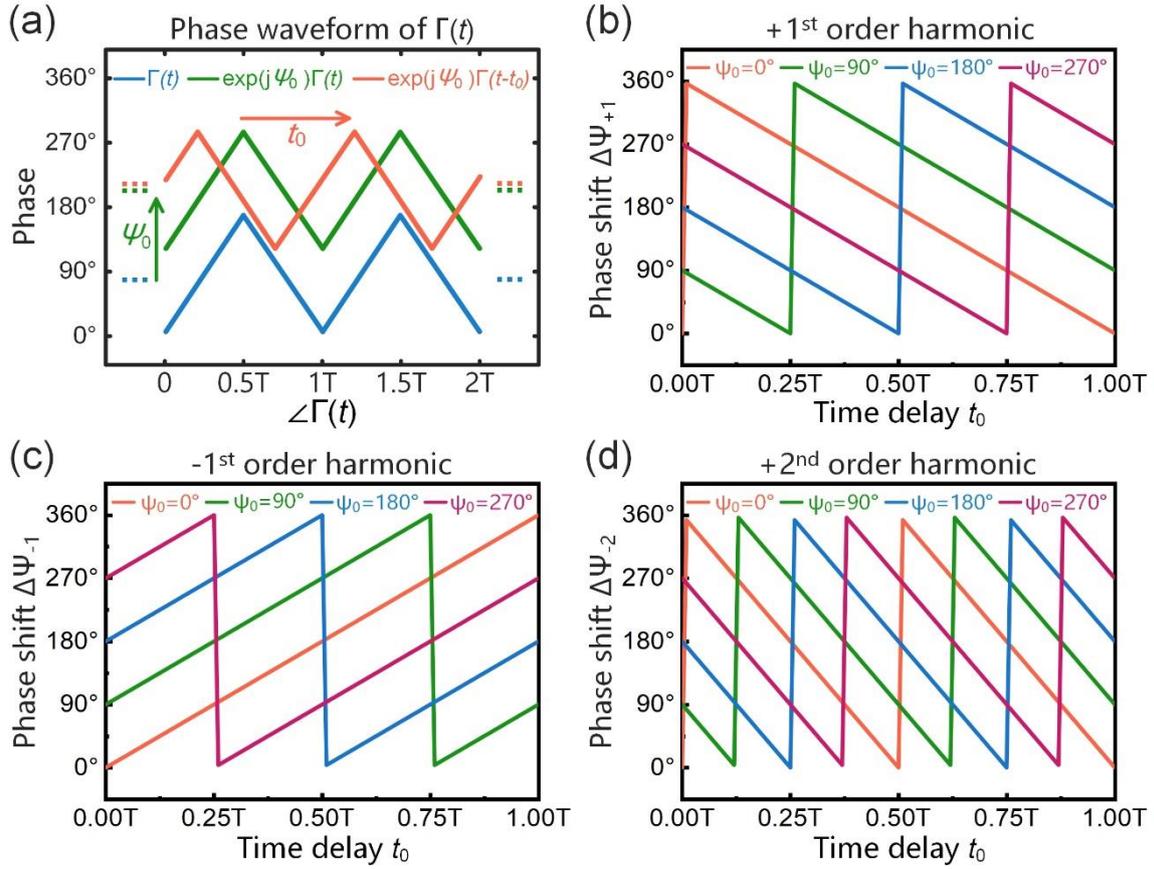

**FIG. 2.** (a) Demonstrations of periodic phase of $\Gamma(t)$, which shows the original waveform (blue line), the waveform with an extra initial phase $\psi_0$ (green line), and the waveform with both $\psi_0$ and time delay $t_0$ (red line), respectively. (b-d) The calculated phase shifts of the $+1^{st}$, $-1^{st}$ and $+2^{nd}$-order harmonics introduced by the multifarious combinations of $\psi_0$ and $t_0$.

Here we refocus on Eq. (4). If we add an extra initial phase $\psi_0$ and introduce a time delay $t_0$ simultaneously to $\Gamma(t)$, as illustrated in Fig. 2(a), then $a_k$ can be recalculated as

$$a_k' = \frac{1}{T}\int_0^T \exp(j\psi_0)\Gamma(t-t_0)\exp(-jk2\pi f_0 t)dt = A_k\exp[j(\Psi_k + \psi_0 - k2\pi f_0 t_0)]. \quad (5)$$

Comparing Eqs. (4) and (5), we note that $\psi_0$ and $t_0$ have similar influence on $a_k$, that is to say,

regulate the phase response while keep the amplitude unchanged. However, the contributions to the phase shift $\Delta\Psi_k$ introduced by $\psi_0$ and $t_0$ are quite different according to the harmonic order. Figures 2(b-d) demonstrate the calculation results of $\Delta\Psi_k$ with multifarious combinations of $\psi_0$ and $t_0$ at different harmonics, respectively. To elucidate this phenomenon analytically, the corresponding phase shifts of the $m^{th}$ and $n^{th}$-order harmonics ($m \neq n$) are listed as

$$\begin{cases} \Delta\Psi_m = \psi_0 - m2\pi f_0 t_0 \\ \Delta\Psi_n = \psi_0 - n2\pi f_0 t_0 \end{cases}. \tag{6}$$

By solving the binary linear equations, the desired $\psi_0$ and $t_0$ are obtained respectively as

$$\begin{cases} \psi_0 = \frac{m\Delta\Psi_n - n\Delta\Psi_m}{(m-n)} \\ t_0 = \frac{\Delta\Psi_n - \Delta\Psi_m}{(m-n)2\pi f_0} \end{cases}. \tag{7}$$

The theoretical derivations clearly demonstrate different effects by $\psi_0$ and $t_0$ on the harmonic phases, which introduce a simple and effective method to implement the phase syntheses of arbitrary dual harmonics simultaneously.

As an intuitive example, we select two harmonics ($+1^{st}$ and $+2^{nd}$) to realize 3-bit coding phases independently, where the eight digits (0, 1, 2, 3, 4, 5, 6, 7) correspond to eight phase states ($0, 0.25\pi, 0.5\pi, 0.75\pi, \pi, 1.25\pi, 1.5\pi, 1.75\pi$). A parameter $\gamma_{t_0}^{\psi_0}$ defined as $\gamma_{t_0}^{\psi_0} = \exp(j\psi_0)\Gamma(t - t_0)$ is introduced to conveniently denote the extra initial phase and time delay added to $\Gamma(t)$. Accordingly, all corresponding 64 combinations of $\psi_0$ and $t_0$ added to $\Gamma(t)$ for dual-harmonic independent 3-bit coding phase syntheses are listed in Table I. To show the robustness, we also calculate the corresponding results for the $+1^{st}$ and $-1^{st}$-order harmonics, which are presented in Table II.

Table I. The desired combinations of $\psi_0$ and $t_0$ added to $\Gamma(t)$ for independent 3-bit coding phase syntheses at $+1^{st}$ and $+2^{nd}$-order harmonics.

| | Digital codes | \multicolumn{8}{c}{$+2^{nd}$-order harmonic} |
|---|---|---|---|---|---|---|---|---|---|
| | | 0 | 1 | 2 | 3 | 4 | 5 | 6 | 7 |
| $+1^{st}$-order harmonic | 0 | $\gamma_0^0$ | $\gamma_{0.875T}^{1.75\pi}$ | $\gamma_{0.75T}^{1.5\pi}$ | $\gamma_{0.625T}^{1.25\pi}$ | $\gamma_{0.5T}^{\pi}$ | $\gamma_{0.375T}^{0.75\pi}$ | $\gamma_{0.25T}^{0.5\pi}$ | $\gamma_{0.125T}^{0.25\pi}$ |
| | 1 | $\gamma_{0.125T}^{0.5\pi}$ | $\gamma_0^{0.25\pi}$ | $\gamma_{0.875T}^{0}$ | $\gamma_{0.75T}^{1.75\pi}$ | $\gamma_{0.625T}^{1.5\pi}$ | $\gamma_{0.5T}^{1.25\pi}$ | $\gamma_{0.375T}^{\pi}$ | $\gamma_{0.25T}^{0.75\pi}$ |
| | 2 | $\gamma_{0.25T}^{\pi}$ | $\gamma_{0.125T}^{0.75\pi}$ | $\gamma_0^{0.5\pi}$ | $\gamma_{0.875T}^{0.25\pi}$ | $\gamma_{0.75T}^{0}$ | $\gamma_{0.625T}^{1.75\pi}$ | $\gamma_{0.5T}^{1.5\pi}$ | $\gamma_{0.375T}^{1.25\pi}$ |
| | 3 | $\gamma_{0.375T}^{1.5\pi}$ | $\gamma_{0.25T}^{1.25\pi}$ | $\gamma_{0.125T}^{\pi}$ | $\gamma_0^{0.75\pi}$ | $\gamma_{0.875T}^{0.5\pi}$ | $\gamma_{0.75T}^{0.25\pi}$ | $\gamma_{0.625T}^{0}$ | $\gamma_{0.5T}^{1.75\pi}$ |
| | 4 | $\gamma_{0.5T}^{0}$ | $\gamma_{0.375T}^{1.75\pi}$ | $\gamma_{0.25T}^{1.5\pi}$ | $\gamma_{0.125T}^{1.25\pi}$ | $\gamma_0^{\pi}$ | $\gamma_{0.875T}^{0.75\pi}$ | $\gamma_{0.75T}^{0.5\pi}$ | $\gamma_{0.625T}^{0.25\pi}$ |
| | 5 | $\gamma_{0.625T}^{0.5\pi}$ | $\gamma_{0.5T}^{0.25\pi}$ | $\gamma_{0.375T}^{0}$ | $\gamma_{0.25T}^{1.75\pi}$ | $\gamma_{0.125T}^{1.5\pi}$ | $\gamma_0^{1.25\pi}$ | $\gamma_{0.875T}^{\pi}$ | $\gamma_{0.75T}^{0.75\pi}$ |
| | 6 | $\gamma_{0.75T}^{\pi}$ | $\gamma_{0.625T}^{0.75\pi}$ | $\gamma_{0.5T}^{0.5\pi}$ | $\gamma_{0.375T}^{0.25\pi}$ | $\gamma_{0.25T}^{0}$ | $\gamma_{0.125T}^{1.75\pi}$ | $\gamma_0^{1.5\pi}$ | $\gamma_{0.875T}^{1.25\pi}$ |
| | 7 | $\gamma_{0.875T}^{1.5\pi}$ | $\gamma_{0.75T}^{1.25\pi}$ | $\gamma_{0.625T}^{\pi}$ | $\gamma_{0.5T}^{0.75\pi}$ | $\gamma_{0.375T}^{0.5\pi}$ | $\gamma_{0.25T}^{0.25\pi}$ | $\gamma_{0.125T}^{0}$ | $\gamma_0^{1.75\pi}$ |

Table II. The desired combinations of $\psi_0$ and $t_0$ added to $\Gamma(t)$ for independent 3-bit coding phase synthesis at $+1^{st}$ and $-1^{st}$-order harmonics.

| | Digital codes | \multicolumn{8}{c}{$-1^{st}$-order harmonic} |
|---|---|---|---|---|---|---|---|---|---|
| | | 0 | 1 | 2 | 3 | 4 | 5 | 6 | 7 |
| $+1^{st}$-order harmonic | 0 | $\gamma_0^0$ | $\gamma_{0.0625T}^{0.125\pi}$ | $\gamma_{0.125T}^{0.25\pi}$ | $\gamma_{0.1875T}^{0.375\pi}$ | $\gamma_{0.25T}^{0.5\pi}$ | $\gamma_{0.3125T}^{0.625\pi}$ | $\gamma_{0.375T}^{0.75\pi}$ | $\gamma_{0.4375T}^{0.875\pi}$ |
| | 1 | $\gamma_{0.9375T}^{0.125\pi}$ | $\gamma_0^{0.25\pi}$ | $\gamma_{0.0625T}^{0.375\pi}$ | $\gamma_{0.125T}^{0.5\pi}$ | $\gamma_{0.1875T}^{0.625\pi}$ | $\gamma_{0.25T}^{0.75\pi}$ | $\gamma_{0.3125T}^{0.875\pi}$ | $\gamma_{0.375T}^{\pi}$ |
| | 2 | $\gamma_{0.875T}^{0.25\pi}$ | $\gamma_{0.9375T}^{0.375\pi}$ | $\gamma_0^{0.5\pi}$ | $\gamma_{0.0625T}^{0.625\pi}$ | $\gamma_{0.125T}^{0.75\pi}$ | $\gamma_{0.1875T}^{0.875\pi}$ | $\gamma_{0.25T}^{\pi}$ | $\gamma_{0.3125T}^{1.125\pi}$ |
| | 3 | $\gamma_{0.8125T}^{0.375\pi}$ | $\gamma_{0.875T}^{0.5\pi}$ | $\gamma_{0.9375T}^{0.625\pi}$ | $\gamma_0^{0.75\pi}$ | $\gamma_{0.0625T}^{0.875\pi}$ | $\gamma_{0.125T}^{\pi}$ | $\gamma_{0.1875T}^{1.125\pi}$ | $\gamma_{0.25T}^{1.25\pi}$ |
| | 4 | $\gamma_{0.75T}^{0.5\pi}$ | $\gamma_{0.8125T}^{0.625\pi}$ | $\gamma_{0.875T}^{0.75\pi}$ | $\gamma_{0.9375T}^{0.875\pi}$ | $\gamma_0^{\pi}$ | $\gamma_{0.0625T}^{1.125\pi}$ | $\gamma_{0.125T}^{1.25\pi}$ | $\gamma_{0.1875T}^{1.375\pi}$ |
| | 5 | $\gamma_{0.6875T}^{0.625\pi}$ | $\gamma_{0.75T}^{0.75\pi}$ | $\gamma_{0.8125T}^{0.875\pi}$ | $\gamma_{0.875T}^{\pi}$ | $\gamma_{0.9375T}^{1.125\pi}$ | $\gamma_0^{1.25\pi}$ | $\gamma_{0.0625T}^{1.375\pi}$ | $\gamma_{0.125T}^{1.5\pi}$ |
| | 6 | $\gamma_{0.625T}^{0.75\pi}$ | $\gamma_{0.6875T}^{0.875\pi}$ | $\gamma_{0.75T}^{\pi}$ | $\gamma_{0.8125T}^{1.125\pi}$ | $\gamma_{0.875T}^{1.25\pi}$ | $\gamma_{0.9375T}^{1.375\pi}$ | $\gamma_0^{1.5\pi}$ | $\gamma_{0.0625T}^{1.625\pi}$ |
| | 7 | $\gamma_{0.5625T}^{0.875\pi}$ | $\gamma_{0.625T}^{\pi}$ | $\gamma_{0.6875T}^{1.125\pi}$ | $\gamma_{0.75T}^{1.25\pi}$ | $\gamma_{0.8125T}^{1.375\pi}$ | $\gamma_{0.875T}^{1.5\pi}$ | $\gamma_{0.9375T}^{1.625\pi}$ | $\gamma_0^{1.75\pi}$ |

**Wavefront manipulations of dual harmonics using the STC digital metasurface.** Based on the synthesis strategy of dual-harmonic phases being fully discussed above, we will step further to investigate the wavefront manipulations of the two harmonics in the space domain. Suppose that each meta-atom owns an independent time-varying reflection coefficient $\Gamma^{pq}(t)$, where the superscript stands for its coordinates on the metasurface. With the help of previous researches on the STC metasurface, the far-field scattering pattern in the time domain can be expressed as [40-42]

$$f(\theta, \varphi, t) = \sum_{p=1}^{M}\sum_{q=1}^{N} E^{pq}(\theta, \varphi)\Gamma^{pq}(t)\exp\left\{j\frac{2\pi}{\lambda_c}[(p-1)d_x\sin\theta\cos\varphi + (q-$$

$1)d_y sin\theta sin\varphi]\}$.        (8)

Here, $E^{pq}(\theta,\varphi) = \cos\theta$ is the scattering pattern of the $(p,q)^{th}$ meta-atom at the central frequency $f_c$, which is approximately a cosine function for simplicity [40, 41], $\theta$ and $\varphi$ are the elevation and azimuth angles, $\lambda_c = c/f_c$ is the wavelength corresponding to the central frequency $f_c$ with $c$ representing the light speed in vacuum, and $d_x$ and $d_y$ are the periods of meta-atom along the *x* and *y* directions, respectively. Through the Fourier series expression of $\Gamma^{pq}(t)$, the far-field scattering pattern at the $k^{th}$-order harmonic frequency $f_c + kf_0$ is given by:

$$F_k(\theta,\varphi) = \sum_{p=1}^{M}\sum_{q=1}^{N} E^{pq}(\theta,\varphi) a_k^{pq} \exp\left\{j\frac{2\pi}{\lambda_c}[(p-1)d_x sin\theta cos\varphi + (q-1)d_y sin\theta sin\varphi]\right\},  \quad (9)$$

where $\lambda_k = c/(f_c + kf_0)$ is the wavelength of the $k^{th}$-order harmonic frequency $f_c + kf_0$.

According to the generalized Snell's law [8], the key to realizing the wavefront manipulation lies in precisely controlling the phase profile of the metasurface, more specifically, the phase gradient of $a_k^{pq}$ (i.e. $\Psi_k^{pq}$) on the metasurface at the $k^{th}$-order harmonic. Combined with Eq. (9), each harmonic can possess an independent $\Psi_k^{pq}$ matrix to control its wavefront propagation. Consequently, it can be shown that the wavefront manipulations of dual harmonics can be achieved as long as the corresponding phase matrices are implemented simultaneously. In order to realize the two matrices, e.g. $\Psi_m^{pq}$ and $\Psi_n^{pq}$, the phase synthesis method proposed in Eq. (7) is a proper candidate, which will finally construct a compact matrix composed of the combinations of extra initial phase and time delay, i.e. $(\gamma_{t_0}^{\psi_0})^{pq}$.

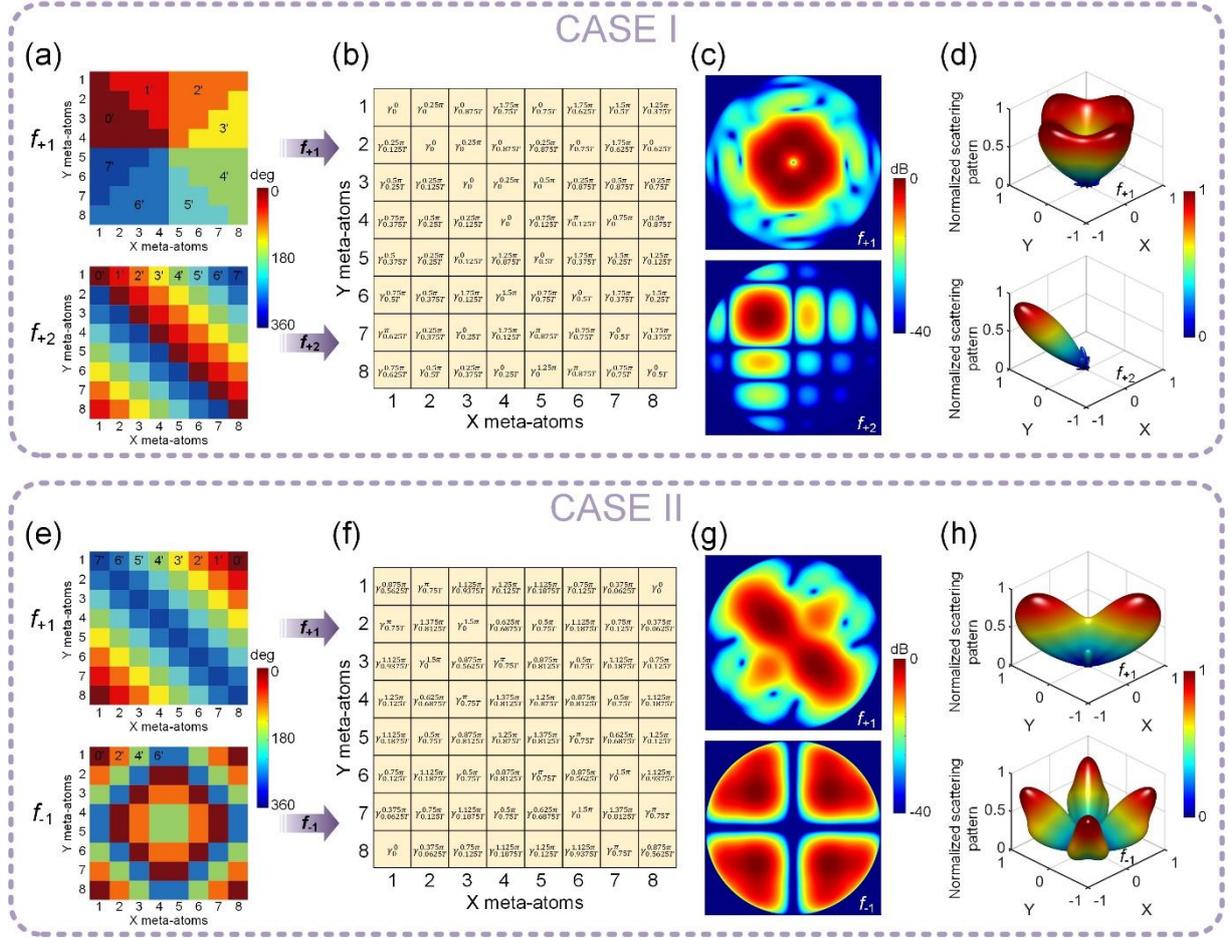

**FIG. 3.** Two examples of dual-harmonic manipulations, where Cases I and II respectively demonstrate the process of independent wavefront shaping for two different pairs of dual-harmonics, namely the $+1^{st}/+2^{nd}$ and $+1^{st}/-1^{st}$-order harmonics. **CASE I:** (a) Two desired phase profiles of dual harmonics using 3-bit coding representation, one of which is a spiral-like distribution for OAM generation (mode $l = 1$) at the $+1^{st}$-order harmonic, and the other is a diagonal gradient distribution for beam steering at the $+2^{nd}$-order harmonic. (b) The calculated compact matrix $(\gamma_{t_0}^{\psi_0})^{pq}$ according to Table I, which is used for simultaneous phase syntheses of the $+1^{st}$ and $+2^{nd}$-order harmonics. (c, d) The corresponding 2D and 3D scattering patterns at the $+1^{st}$ and $+2^{nd}$-order harmonics, respectively. **CASE II:** Similar configuration to that in **CASE I**. (e) Two novel phase profiles for rabbit-ear double beams at the $+1^{st}$-order harmonic and symmetrically four split beams at the $-1^{st}$-order harmonic. (f) The newly calculated compact matrix $(\gamma_{t_0}^{\psi_0})^{pq}$ according to Table II. (g, h) The corresponding 2D and 3D scattering patterns at the $+1^{st}$ and $-1^{st}$-order harmonics, respectively.

For more intuitive demonstration, we study two cases with the help of Tables I and II to show independent wavefront manipulations of dual harmonics. We assume that the STC digital metasurface consists of $8 \times 8$ meta-atoms with the following parameters: $f_c = 5GHz$, $f_0 =$ $100 kHz$, and $d_x = d_y = \lambda_c/3$. In the first case, as shown in Fig. 3(a), we exhibit a spiral-like phase profile at the $+1^{st}$-order harmonic and a diagonal gradient profile at the $+2^{nd}$-order harmonic. Both phase profiles are transformed into two $8 \times 8$ matrices, i.e. $\Psi_{+1}^{pq}$ and $\Psi_{+2}^{pq}$, where the digits from '0' to '7' refer to the eight phase states respectively under the 3-bit coding condition. According to Table I, the compact matrix $(\gamma_{t_0}^{\psi_0})^{pq}$ can be calculated to synthesize the two phase matrices $\Psi_{+1}^{pq}$ and $\Psi_{+2}^{pq}$ simultaneously and the results are illustrated in Fig. 3(b). The corresponding 2D and 3D scattering patterns at the $+1^{st}$ and $+2^{nd}$-order harmonics are respectively shown in Figs. 3(c) and 3(d), which demonstrate a vortex beam carrying the OAM mode for the $+1^{st}$-order harmonic and beam steering to a specific angle for the $+2^{nd}$-order harmonic. For Case II shown in Figs. 3(e-h), similar process is implemented when considering the $+1^{st}$ and $-1^{st}$-order harmonic manipulations simultaneously. Here, the reference table is changed to Table II before calculating the final compact matrix $(\gamma_{t_0}^{\psi_0})^{pq}$. Two different beam shapes, namely the rabbit-ear double beams and symmetrically four split beams, are generated at the $+1^{st}$ and $-1^{st}$-order harmonics, respectively, as shown in Fig. 3(h).

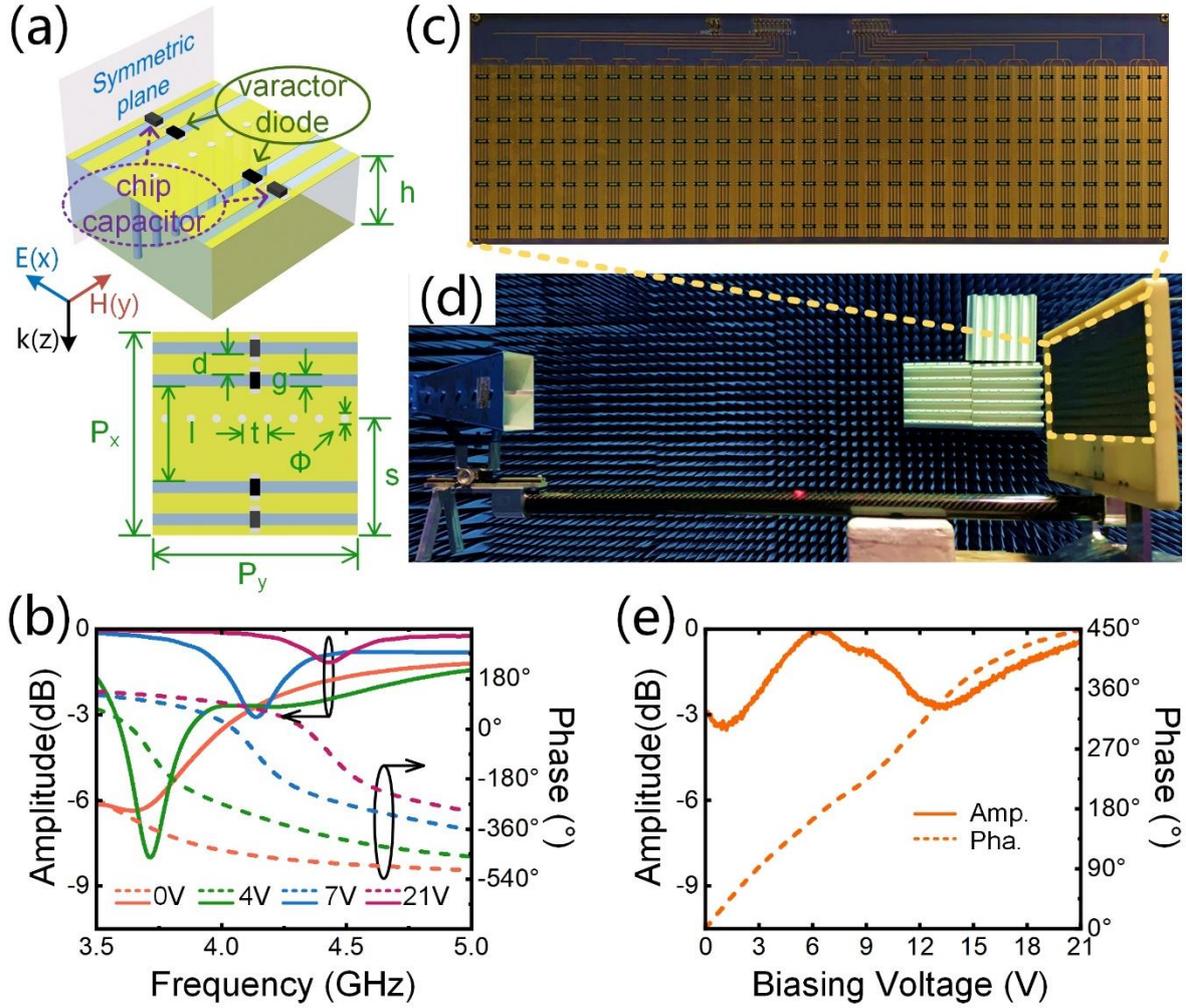

**FIG. 4.** (a) The details and geometric parameters of half meta-atom with the other half hidden behind the symmetric plane. (b) The simulated reflection amplitude and phase responses of the meta-atom under different biasing voltages. (c, d) The photographs of the STC metasurface sample and the experiment environment in the anechoic chamber. (e) The measured reflection amplitude and phase responses of the STC metasurface under the illumination of EM wave at 4.25GHz with different biasing voltages from 0V to 21V.

## III. DESIGN AND EXPERIMENTS

**Design of the STC digital metasurface.** According to previous derivations, the proposed phase synthesis method could be used to the maximum extent if the STC metasurface possesses a full range of phase tunability. Here, we use the metasurface in Ref. [30] to validate the proposed theory and method. The geometry of half meta-atom is firstly illustrated in Fig. 4(a), where the corresponding parameters are $P_x$ = 12 mm, $P_y$ = 12 mm, h = 5 mm, s = 6.9 mm, l = 5.6 mm, g = 0.7 mm, d = 1.2 mm, and t = 1.5 mm, and the via hole diameter is $\Phi$ = 0.4 mm. Several metallic strips with different widths are printed on a 5mm-thick grounded F4B substrate ($\varepsilon_r$ = 3.0 and tan $\sigma$ = 0.0015), with two varactor diodes (SMV-2019, Skyworks, Inc.) and two chip capacitors (0.1pF) mounted between different strips. The widest metallic strip is connected to the ground through a row of via holes along the *x* direction, constructing two different cavities to achieve large phase tuning range. The other half of the meta-atom is the mirror image of the displayed structure according to the symmetric plane. Full-wave simulations are performed by using the commercial software, CST Microwave Studio 2016, to characterize the reflection responses of meta-atom under different biasing voltages. In simulations, periodic boundaries are applied in the *x* and *y* directions to mimic an infinite array, and Floquet port is set on the top of the element to emit a plane wave polarized in the *x* direction to the meta-atom. In addition, RLC series circuit is selected to model the varactor diode, in which the equivalent parameters under different biasing voltages are listed in Ref. [42]. In Fig. 4(b), we display the simulated reflection spectra of the meta-atom under different biasing voltages from 0 to 21V. The results show a large phase tuning range over 500° around 4.25GHz, and the corresponding amplitude fluctuation is less than 3dB, which confirms the feasibility of the designed meta-atom as the STC metasurface elements for phase syntheses of dual harmonics.

**Fabrication.** Based on the proposed design, an STC digital metasurface composed of 16 × 8 meta-atoms was fabricated in Ref. [30] using the printed circuit board (PCB) technology. Figure 4(c) shows the photograph of the sample, which occupies 384×126 mm$^2$ including the biasing network. Each column of the meta-atoms shares the same control signal, making the STC metasurface a 16-column linear array. According to our previous work [30], the measured reflection spectra of the STC metasurface at 4.25GHz via different biasing voltages are illustrated in Fig. 4(e). The results show that, as the biasing voltage changes from 0 to 21V, the STC metasurface gains a large phase tuning range of 2.5$\pi$ with an amplitude fluctuation less than 3.5dB, which have good agreements with the simulation results. The slight deviations are reasonable due to the imperfection of devices, inaccuracy of parameters, and tolerance of fabrication. Hence the STC digital metasurface is an ideal candidate for implementing the manipulation theory of dual-harmonic waves.

**Experiments and results.** In our experiments, the monochromic signal is generated by a microwave signal generator (Agilent E8257D) and then emitted through a horn antenna to illuminate the STC digital metasurface. As shown in Fig. 4(d), a simple fixture is used to locate and clamp the metasurface sample and the feed antenna, which are secured tightly on a mechanical turntable. At the receiving terminal, another horn antenna connected to a spectrum analyzer (Agilent EE4447A) is employed to capture the reflected wave and measure its spectral intensity. A customized platform is used to generate the periodic time-varying signals, which are converted into biasing voltages and then applied to the STC metasurface. The operating frequency $f_c$ is 4.25GHz and the detailed configuration of the platform is illustrated in Table III.

Table III. Detailed configuration of the customized platform.

| Type | Model | Features |
| --- | --- | --- |
| Controller | PXIe-8135 | Intel Core i7-3610QE quad-core processor, 2.3 GHz base frequency |
| Chassis | PXIe-1082 | 8 GB/s bus bandwidth PXIe chassis with 8 slots |
| FPGA Module | PXIe-7976R | Kintex-7 410T FPGA, 2 GB DRAM |
| DAC Module | NI-5783 | Analog four input/output FlexRIO adapter module, 100 MS/s sample rate, 400 MS/s update rate |
| DC Power Supply | PXI-4110 | Programmable DC power supply, ± 20 volts voltage range |
| Timing Module | PXIe-6674T | 10 MHz clock based on an onboard precision OCXO reference |

Before measurements, we need to establish a basic waveform of $\Gamma(t)$ to realize harmonic generation. Although this is not the focus of this work, it is a prerequisite indeed. Considering the EM characteristics of our STC metasurface, for simplicity, we use a square wave with 25% duty cycle and period of 10μs as the basic waveform, which can be referred to supplementary material S1 for detailed information. This kind of waveform possesses relatively abundant harmonic components, e.g. the aforementioned $-1^{st}$, $+1^{st}$, and $+2^{nd}$-order harmonics, which make the effect of dual-harmonic manipulations more obvious. Then we redesign the waveform of $\Gamma(t)$ by adding various combinations of $\psi_0$ and $t_0$ from Tables I and II, respectively, to realize dual-phase syntheses for the corresponding harmonics. Considering that our platform is equipped with only 4 individual digital-analog convertors (DACs), we use 2-bit coding strategy for experimental validation, which contains four digits (0, 1, 2, 3) to stand for four phase states $(0, 0.5\pi, \pi, 1.5\pi)$ respectively. The detailed waveform configurations of the 2-bit coding are given in supplementary material S2. Finally, combined with the realistic voltage-phase curve shown in Fig. 4(e), we programed the control platform to generate the corresponding biasing signals to obtain the desired waveforms.

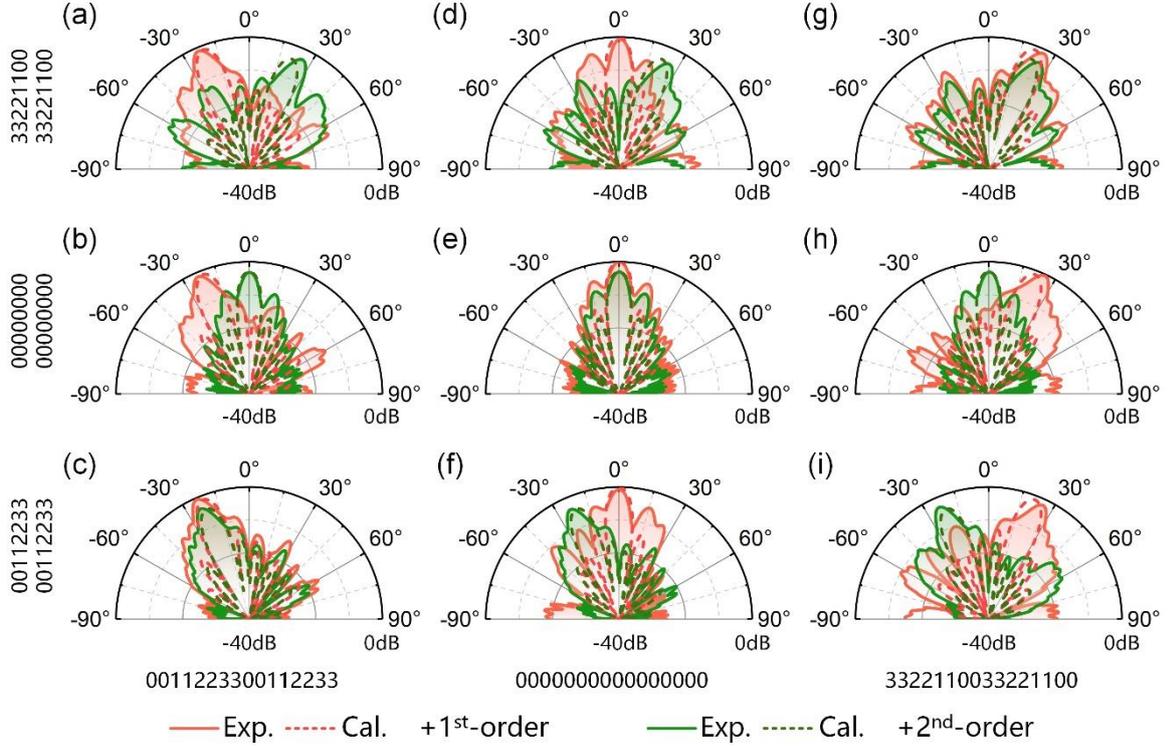

**FIG. 5.** The normalized calculation (dashed line) and experimental (solid line) results of the $+1^{st}$ (red lines) and $+2^{nd}$ (green lines) harmonics, which operate at 4.2501GHz and 4.2502GHz, respectively. Three sets of 2-bit coding sequences, '0011223300112233', '0000000000000000', and '3322110033221100', are chosen as the candidates for wavefront manipulations. Each column from the left to the right (a-c, d-f, g-i) shares the same coding sequence shown at the bottom to manipulate the scattering patterns of the $+1^{st}$-order harmonic, while each row from the top to bottom (a-g, b-h, c-i) demonstrates the corresponding results of the $+2^{nd}$-order harmonic.

Table IV. All configurations of compact matrix $(\gamma_{t_0}^{\psi_0})^{pq}$ for implementing different coding sequences to the $+1^{st}$ and $+2^{nd}$-order harmonics, respectively.

| Coding sequences | | $+2^{nd}$-order harmonic | | |
|---|---|---|---|---|
| | | 0011223300112233 | 0000000000000000 | 3322110033221100 |
| $+1^{st}$-order harmonic | 0011223300112233 | $\gamma_0^0\gamma_0^0\gamma_0^{0.5\pi}\gamma_0^{0.5\pi}$ $\gamma_0^\pi\gamma_0^\pi\gamma_0^{1.5\pi}\gamma_0^{1.5\pi}$ $\gamma_0^0\gamma_0^0\gamma_0^{0.5\pi}\gamma_0^{0.5\pi}$ $\gamma_0^\pi\gamma_0^\pi\gamma_0^{1.5\pi}\gamma_0^{1.5\pi}$ | $\gamma_0^0\gamma_0^0\gamma_{0.25T}^\pi\gamma_{0.25T}^\pi$ $\gamma_{0.5T}^0\gamma_{0.5T}^0\gamma_{0.75T}^\pi\gamma_{0.75T}^\pi$ $\gamma_0^0\gamma_0^0\gamma_{0.25T}^\pi\gamma_{0.25T}^\pi$ $\gamma_{0.5T}^0\gamma_{0.5T}^0\gamma_{0.75T}^\pi\gamma_{0.75T}^\pi$ | $\gamma_{0.25T}^{0.5\pi}\gamma_{0.25T}^{0.5\pi}\gamma_{0.75T}^0\gamma_{0.75T}^0$ $\gamma_{0.25T}^{1.5\pi}\gamma_{0.25T}^{1.5\pi}\gamma_{0.75T}^\pi\gamma_{0.75T}^\pi$ $\gamma_{0.25T}^{0.5\pi}\gamma_{0.25T}^{0.5\pi}\gamma_{0.75T}^0\gamma_{0.75T}^0$ $\gamma_{0.25T}^{1.5\pi}\gamma_{0.25T}^{1.5\pi}\gamma_{0.75T}^\pi\gamma_{0.75T}^\pi$ |
| | 0000000000000000 | $\gamma_0^0\gamma_0^0\gamma_{0.75T}^{1.5\pi}\gamma_{0.75T}^{1.5\pi}$ $\gamma_{0.5T}^\pi\gamma_{0.5T}^\pi\gamma_{0.25T}^{0.5\pi}\gamma_{0.25T}^{0.5\pi}$ $\gamma_0^0\gamma_0^0\gamma_{0.75T}^{1.5\pi}\gamma_{0.75T}^{1.5\pi}$ $\gamma_{0.5T}^\pi\gamma_{0.5T}^\pi\gamma_{0.25T}^{0.5\pi}\gamma_{0.25T}^{0.5\pi}$ | $\gamma_0^0\gamma_0^0\gamma_0^0\gamma_0^0$ $\gamma_0^0\gamma_0^0\gamma_0^0\gamma_0^0$ $\gamma_0^0\gamma_0^0\gamma_0^0\gamma_0^0$ $\gamma_0^0\gamma_0^0\gamma_0^0\gamma_0^0$ | $\gamma_{0.25T}^{0.5\pi}\gamma_{0.25T}^{0.5\pi}\gamma_{0.5T}^\pi\gamma_{0.5T}^\pi$ $\gamma_{0.75T}^{1.5\pi}\gamma_{0.75T}^{1.5\pi}\gamma_0^0\gamma_0^0$ $\gamma_{0.25T}^{0.5\pi}\gamma_{0.25T}^{0.5\pi}\gamma_{0.5T}^\pi\gamma_{0.5T}^\pi$ $\gamma_{0.75T}^{1.5\pi}\gamma_{0.75T}^{1.5\pi}\gamma_0^0\gamma_0^0$ |
| | 3322110033221100 | $\gamma_{0.75T}^\pi\gamma_{0.75T}^\pi\gamma_{0.25T}^{1.5\pi}\gamma_{0.25T}^{1.5\pi}$ $\gamma_{0.75T}^0\gamma_{0.75T}^0\gamma_{0.25T}^{0.5\pi}\gamma_{0.25T}^{0.5\pi}$ $\gamma_{0.75T}^\pi\gamma_{0.75T}^\pi\gamma_{0.25T}^{1.5\pi}\gamma_{0.25T}^{1.5\pi}$ $\gamma_{0.75T}^0\gamma_{0.75T}^0\gamma_{0.25T}^{0.5\pi}\gamma_{0.25T}^{0.5\pi}$ | $\gamma_{0.75T}^\pi\gamma_{0.75T}^\pi\gamma_{0.5T}^0\gamma_{0.5T}^0$ $\gamma_{0.25T}^\pi\gamma_{0.25T}^\pi\gamma_0^0\gamma_0^0$ $\gamma_{0.75T}^\pi\gamma_{0.75T}^\pi\gamma_{0.5T}^0\gamma_{0.5T}^0$ $\gamma_{0.25T}^\pi\gamma_{0.25T}^\pi\gamma_0^0\gamma_0^0$ | $\gamma_0^{1.5\pi}\gamma_0^{1.5\pi}\gamma_0^\pi\gamma_0^\pi$ $\gamma_0^{0.5\pi}\gamma_0^{0.5\pi}\gamma_0^0\gamma_0^0$ $\gamma_0^{1.5\pi}\gamma_0^{1.5\pi}\gamma_0^\pi\gamma_0^\pi$ $\gamma_0^{0.5\pi}\gamma_0^{0.5\pi}\gamma_0^0\gamma_0^0$ |

We firstly investigate the simultaneous manipulations of the scattering patterns for the $+1^{st}$ and $+2^{nd}$-order harmonics via the STC metasurface. The period of $\Gamma(t)$ is set to 10μs, indicating that the $+1^{st}$ and $+2^{nd}$-order harmonics of the reflected wave operate at frequencies 4.2501 and 4.2502 GHz, respectively. As the STC metasurface contains 16 columns of meta-atoms, we choose three sets of coding sequences, '0011223300112233', '0000000000000000' and '3322110033221100', as the candidates for the two phase matrices $\Psi_{+1}^{pq}$ and $\Psi_{+2}^{pq}$. Then, as listed in Table IV, there are 9 different configurations of compact matrix $(\gamma_{t_0}^{\psi_0})^{pq}$ in total to manipulate these two harmonics simultaneously. Figs. 5(a-i) demonstrate the measured scattering patterns of the $+1^{st}$ and $+2^{nd}$-order harmonics, respectively, with different compact matrices employed in the STC metasurface, along with the corresponding calculation results for comparisons. Each subfigure in Fig. 5 denotes the scattering patterns of two harmonics under one configuration of $(\gamma_{t_0}^{\psi_0})^{pq}$. Here, three columns of subfigures from the left to right (Figs. 5(a-c), 5(d-f), 5(g-i)) respectively stand for the scattering patterns of the $+1^{st}$-order harmonic under the three coding sequences; while three rows of subfigures from the top to bottom (Figs. 5(a-g), 5(b-h), 5(c-i)) are the scattering patterns of the $+2^{nd}$-order harmonic under the same coding sequences. These results clearly show how the coding sequences affect the scattering patterns of the two harmonics based on the generalized Snell's law.

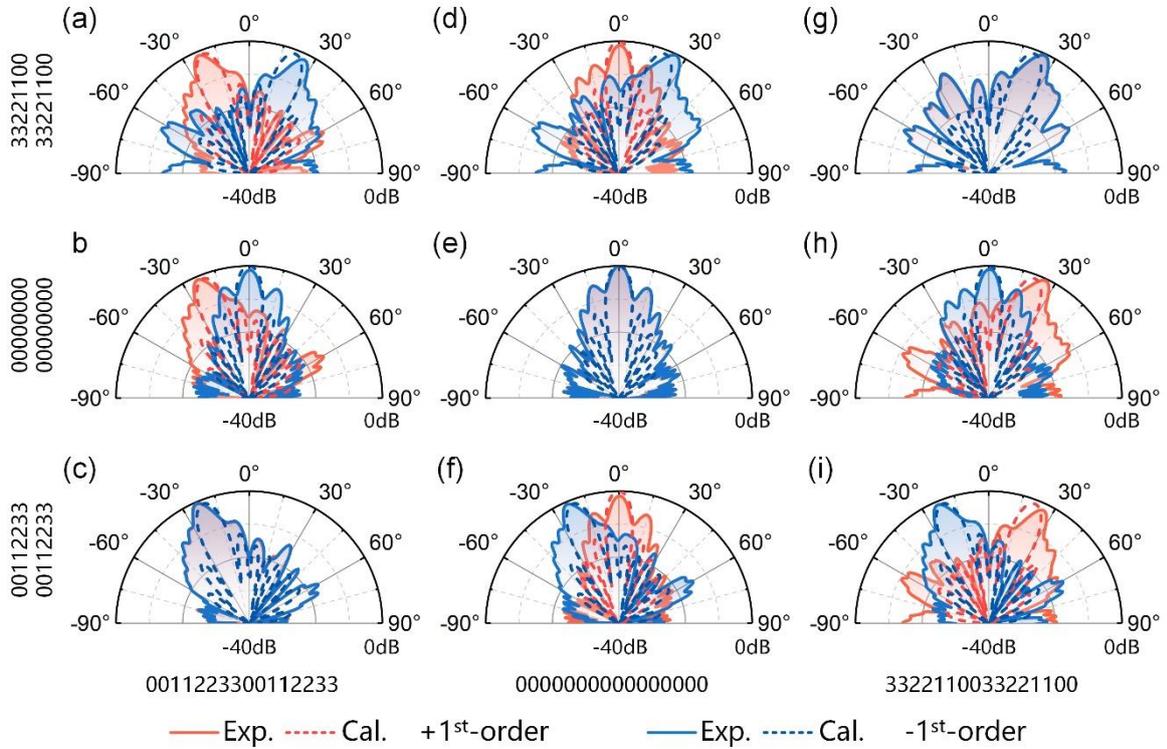

**FIG. 6.** The normalized calculation (dashed line) and experimental (solid line) results of $+1^{st}$ (red line) and $-1^{st}$ (blue line)-order harmonics, which operate at 4.2501GHz and 4.2499GHz, respectively. The results are organized the same as Fig. 5.

Table V. All configurations of compact matrix $(\gamma_{t_0}^{\psi_0})^{pq}$ for implementing different coding sequences to the $+1^{st}$ and $-1^{st}$-order harmonics, respectively.

|  |  | $-1^{st}$-order harmonic | | |
|---|---|---|---|---|
|  |  | 0011223300112233 | 0000000000000000 | 3322110033221100 |
| $+1^{st}$-order harmonic | 0011223300112233 | $\gamma_0^0\gamma_0^0\gamma_0^{0.5\pi}\gamma_0^{0.5\pi}$ <br> $\gamma_0^\pi\gamma_0^\pi\gamma_0^{1.5\pi}\gamma_0^{1.5\pi}$ <br> $\gamma_0^0\gamma_0^0\gamma_0^{0.5\pi}\gamma_0^{0.5\pi}$ <br> $\gamma_0^\pi\gamma_0^\pi\gamma_0^{1.5\pi}\gamma_0^{1.5\pi}$ | $\gamma_0^0\gamma_0^0\gamma_{0.875T}^{0.25\pi}\gamma_{0.875T}^{0.25\pi}$ <br> $\gamma_{0.75T}^{0.5\pi}\gamma_{0.75T}^{0.5\pi}\gamma_{0.625T}^{0.75\pi}\gamma_{0.625T}^{0.75\pi}$ <br> $\gamma_0^0\gamma_0^0\gamma_{0.875T}^{0.25\pi}\gamma_{0.875T}^{0.25\pi}$ <br> $\gamma_{0.75T}^{0.5\pi}\gamma_{0.75T}^{0.5\pi}\gamma_{0.625T}^{0.75\pi}\gamma_{0.625T}^{0.75\pi}$ | $\gamma_{0.375T}^{0.75\pi}\gamma_{0.375T}^{0.75\pi}\gamma_{0.125T}^{0.75\pi}\gamma_{0.125T}^{0.75\pi}$ <br> $\gamma_{0.875T}^{0.75\pi}\gamma_{0.875T}^{0.75\pi}\gamma_{0.625T}^{0.75\pi}\gamma_{0.625T}^{0.75\pi}$ <br> $\gamma_{0.375T}^{0.75\pi}\gamma_{0.375T}^{0.75\pi}\gamma_{0.125T}^{0.75\pi}\gamma_{0.125T}^{0.75\pi}$ <br> $\gamma_{0.875T}^{0.75\pi}\gamma_{0.875T}^{0.75\pi}\gamma_{0.625T}^{0.75\pi}\gamma_{0.625T}^{0.75\pi}$ |
| | 0000000000000000 | $\gamma_0^0\gamma_0^0\gamma_{0.125T}^{0.25\pi}\gamma_{0.125T}^{0.25\pi}$ <br> $\gamma_{0.25T}^{0.5\pi}\gamma_{0.25T}^{0.5\pi}\gamma_{0.375T}^{0.75\pi}\gamma_{0.375T}^{0.75\pi}$ <br> $\gamma_0^0\gamma_0^0\gamma_{0.125T}^{0.25\pi}\gamma_{0.125T}^{0.25\pi}$ <br> $\gamma_{0.25T}^{0.5\pi}\gamma_{0.25T}^{0.5\pi}\gamma_{0.375T}^{0.75\pi}\gamma_{0.375T}^{0.75\pi}$ | $\gamma_0^0\gamma_0^0\gamma_0^0\gamma_0^0$ <br> $\gamma_0^0\gamma_0^0\gamma_0^0\gamma_0^0$ <br> $\gamma_0^0\gamma_0^0\gamma_0^0\gamma_0^0$ <br> $\gamma_0^0\gamma_0^0\gamma_0^0\gamma_0^0$ | $\gamma_{0.375T}^{0.75\pi}\gamma_{0.375T}^{0.75\pi}\gamma_{0.25T}^{0.5\pi}\gamma_{0.25T}^{0.5\pi}$ <br> $\gamma_{0.125T}^{0.25\pi}\gamma_{0.125T}^{0.25\pi}\gamma_0^0\gamma_0^0$ <br> $\gamma_{0.375T}^{0.75\pi}\gamma_{0.375T}^{0.75\pi}\gamma_{0.25T}^{0.5\pi}\gamma_{0.25T}^{0.5\pi}$ <br> $\gamma_{0.125T}^{0.25\pi}\gamma_{0.125T}^{0.25\pi}\gamma_0^0\gamma_0^0$ |
| | 3322110033221100 | $\gamma_{0.625T}^{0.75\pi}\gamma_{0.625T}^{0.75\pi}\gamma_{0.875T}^{0.75\pi}\gamma_{0.875T}^{0.75\pi}$ <br> $\gamma_{0.125T}^{0.75\pi}\gamma_{0.125T}^{0.75\pi}\gamma_{0.375T}^{0.75\pi}\gamma_{0.375T}^{0.75\pi}$ <br> $\gamma_{0.625T}^{0.75\pi}\gamma_{0.625T}^{0.75\pi}\gamma_{0.875T}^{0.75\pi}\gamma_{0.875T}^{0.75\pi}$ <br> $\gamma_{0.125T}^{0.75\pi}\gamma_{0.125T}^{0.75\pi}\gamma_{0.375T}^{0.75\pi}\gamma_{0.375T}^{0.75\pi}$ | $\gamma_{0.625T}^{0.75\pi}\gamma_{0.625T}^{0.75\pi}\gamma_{0.75T}^{0.5\pi}\gamma_{0.75T}^{0.5\pi}$ <br> $\gamma_{0.875T}^{0.25\pi}\gamma_{0.875T}^{0.25\pi}\gamma_0^0\gamma_0^0$ <br> $\gamma_{0.625T}^{0.75\pi}\gamma_{0.625T}^{0.75\pi}\gamma_{0.75T}^{0.5\pi}\gamma_{0.75T}^{0.5\pi}$ <br> $\gamma_{0.875T}^{0.25\pi}\gamma_{0.875T}^{0.25\pi}\gamma_0^0\gamma_0^0$ | $\gamma_0^{1.5\pi}\gamma_0^{1.5\pi}\gamma_0^\pi\gamma_0^\pi$ <br> $\gamma_0^{0.5\pi}\gamma_0^{0.5\pi}\gamma_0^0\gamma_0^0$ <br> $\gamma_0^{1.5\pi}\gamma_0^{1.5\pi}\gamma_0^\pi\gamma_0^\pi$ <br> $\gamma_0^{0.5\pi}\gamma_0^{0.5\pi}\gamma_0^0\gamma_0^0$ |

Next, we conduct another experiment of manipulating the scattering patterns for the $+1^{st}$ and $-1^{st}$-order harmonics via the same STC metasurface to demonstrate the robustness of the proposed approach. The experimental setup remains the same, hence the operation frequencies of the $+1^{st}$ and $-1^{st}$ harmonics are 4.2501 and 4.2499 GHz, respectively. Meanwhile, the same sets of coding sequences are used for the new phase matrices $\Psi_{+1}^{pq}$ and $\Psi_{-1}^{pq}$. Another 9 configurations of the compact matrix $(\gamma_{t_0}^{\psi_0})^{pq}$ are listed in Table V. Finally, we achieve very similar measurements for the $+1^{st}$ and $-1^{st}$-order harmonics, which are respectively shown in Fig. 6. As before, the calculated scattering patterns of the two harmonics (dashed lines) are also displayed for comparison.

From Figs. 5 and 6, it is obvious that the STC digital metasurface successfully realizes simultaneous wavefront manipulations of dual harmonics. By comparing the two experimental results, we note that the scattering patterns of the $+2^{nd}$-order harmonic are about 3 dB lower than those of the $+1^{st}$-order harmonic when employing the same coding sequences. However, the scattering patterns of the $+1^{st}$ and $-1^{st}$-order harmonics are almost equal under the same conditions. Such interesting phenomenon is mostly caused by the spectral distribution of the basic harmonic generation function. Theoretically, there is a 3 dB amplitude attenuation in the $+1^{st}$ and $+2^{nd}$-order harmonics, while the amplitudes of the $+1^{st}$ and $-1^{st}$-order harmonics are always equal due to the symmetrical spectrum distribution (See supplementary material S1 for detailed information). Therefore, it is reasonable to infer that the proposed method has little influence on the energy distribution of harmonics. This unique feature is quite helpful to separate the progress of harmonic generation and wavefront manipulation, thereby simplifying the design process significantly. We also note that the measured scattering patterns have good consistency to the calculated results in different sets of dual harmonics, which validates the theoretical derivations with sufficient accuracy and robustness. Evidently, the proposed STC strategy is more flexible and effective in harmonic manipulations, which is promising for the modern systems such as radar and wireless communication.

## IV. CONCLUSION

In summary, a novel STC strategy was proposed to synthesize multi-bit phases and manipulate wavefronts for arbitrary dual harmonics. Different from previous work of the STC metasurfaces that focus on generating and regulating one specific harmonic, both an initial phase and a time delay are employed in the periodic time-varying phase function to realize the flexible controls of two selected harmonics simultaneously. The novel idea has been analytically proved from theoretical analyses and two sets of dual-harmonic pairs were investigated and simulated as examples. Finally, a proof-of-concept system based on the STC metasurface was implemented for experimental validation. Overall, the proposed approach paves a simple and effective pathway towards designing multifunctional programmable meta-devices, which may bring huge potentials in holographic imaging, intelligent antenna, cognitive radar, and multi-user wireless communication systems. By employing advanced modulation technologies [35], the proposed work can be further extended into the terahertz, optical and acoustic regimes.

## AUTHORS' CONTRIBUTION

J. Y. Dai and J. Yang contributed equally to this work.